\documentclass[a4paper,11pt]{article}
\pdfoutput=1 

\usepackage{jheppub} 
\newcommand{\OO}{\mathcal{O}}
\def\beq#1\eeq{\begin{align}#1\end{align}}
\newcommand{\sut}{$SU(3)$}
\newcommand{\uone}{$U(1)$}
\newcommand{\eps}{\varepsilon}
\newcommand{\system}[1]{\left\{\begin{array}{l} #1\end{array}\right.}

\newcommand{\dbar}{$\overline{\text{d}}$}

\def \mt {\tilde m}
\def \mtr { m_{3/2}}
\def \l {\lambda}
\def \lijk {\lambda''_{ijk}}
\newcommand{\fru}[2]{\left( \frac{#1}{\text{#2}}\right)}

\newcommand{\frud}[2]{\left( \frac{#1}{{#2}}\right)}
\newcommand{\cbs}{$cbs$}
\newcommand{\uds}{$uds$}
\newcommand{\tbs}{$tbs$}

\def \tb {\tan\beta}
\def \tev {\text{ TeV}}
\def \gev {\text{ GeV}}
\def \antid {antideuteron}

\newcommand{\rpv}{$R$-parity violation}
\newcommand{\rpa}{$R$-parity}
\newcommand{\rpcing}{$R$-parity conserving\ }
\newcommand{\rpving}{$R$-parity violating\ }

\title{Gravitino Dark Matter and Flavor Symmetries}

\author{Angelo Monteux,}
\author{Eric Carlson and}
\author{Jonathan M. Cornell}
\affiliation{Santa Cruz Institute for Particle Physics and\\
 Department of Physics, 
  University of California, Santa Cruz CA 95064}

\emailAdd{amonteux@ucsc.edu}
\emailAdd{erccarls@ucsc.edu}
\emailAdd{jcornell@ucsc.edu}
\preprint{SCIPP 14/09}

\abstract{
In supersymmetric theories without $R$-parity, the gravitino can play the role of a decaying Dark Matter candidate without  the problem of late NLSP decays affecting Big Bang Nucleosynthesis. 
In this work, we elaborate on recently discussed limits on $R$-parity violating couplings from decays to antideuterons and discuss the implications for two classes of flavor symmetries: horizontal symmetries, and Minimal Flavor Violation. In most of the parameter space the antideuteron constraints on $R$-parity violating couplings are stronger than low-energy baryon-number-violating processes. Even in the absence of flavor symmetries, we find strong new limits on couplings involving third-generation fields, and discuss the implications for LHC phenomenology. For TeV scale superpartners, we find that the allowed MFV parameter space is a corner with gravitino masses smaller than $\OO(10)$ GeV and small $\tan\beta$.
}

\begin{document} 
\maketitle
\flushbottom

\section{Introduction}

In supersymmetric theories, \rpa \cite{Farrar:1978xj,Bento:1987mu} is usually introduced to remove unwanted dimension four operators that would lead to fast proton decay; the renormalizable \rpving superpotential is:
\beq
W_{RPV}=\mu_i L_i\phi_u+\lambda_{ijk} L_iL_j\bar\ell_k +\lambda'_{ijk}L_iQ_j\bar d_k+\lambda_{ijk}''\bar u_i\bar d_j\bar d_k\,,
\eeq
where the indices are generation indices, $i,j,k=1,\ldots,3$, and only antisymmetric combinations of $i,j$ (respectively, $j,k$) are allowed in $\lambda$ (respectively, $\lambda''$). The first three operators violate lepton number while the last violates baryon number, and both types of operators are involved in proton decay. It is then possible for the proton to be stable if only one type of operators is allowed, leaving $B$ (or $L$) as an accidental symmetry of the theory \cite{Ibanez:1994ig,Dreiner:2005rd}.

This is an aspect of the flavor problems associated with low energy Supersymmetry (SUSY): generic soft terms give large contributions to flavor-changing neutral currents (FCNCs), which can be suppressed by assuming that flavor symmetries govern the structure of the Minimal Supersymmetric Standard Model (MSSM) Lagrangian.
Two particularly well motivated types of flavor symmetries are Abelian horizontal symmetries ({\it a la} Froggatt-Nielsen \cite{Froggatt:1978nt,Leurer:1992wg,Nir:1993mx}) and Minimal Flavor Violation (MFV) \cite{Nikolidakis:2007fc,Csaki:2011ge}, according to which the Higgs Yukawa operators are spurions of a $\sut^5$ flavor symmetry under which the full MSSM Lagrangian is invariant.

Under the assumption of these flavor symmetries, definite  structures of the RPV couplings are predicted:
\begin{itemize}
\item with a horizontal \uone\ symmetry, the relative structure of the RPV couplings is completely determined by the fermion masses and mixings alone \cite{Monteux:2013mna,Joshipura:2000sn,Florez:2013mxa,Sierra:2009zq}; the baryon number violating (BNV) or lepton number violating (LNV) operators are allowed or forbidden independently. In \cite{Monteux:2013mna}, it was argued that, in order not to disagree with LHC null results, LNV operators should be forbidden altogether when considering sub-TeV SUSY. The BNV couplings $\lambda''_{ijk}$ are written in terms of 
an overall scale $\lambda''_{323}
$ and depend on the horizontal charges (we denote the charge of a field by the field symbol itself, $\Phi\equiv q_\Phi$, and the inter-generational difference between two fields as $\Phi_{ij}\equiv q_{\Phi_i}-q_{\Phi_j}$):
\beq
&\lijk=\l''_{323}\eps^{{u_{i3}}+d_{j2}+d_{k3}}, \qquad \left(\text{where }\eps\equiv V^{CKM}_{12}\simeq \sin\theta_C\right),
\\\label{LHOR}
&\qquad\left(\begin{array}{ccc} \lambda''_{112}  & \lambda''_{212}  & \lambda''_{312}  \\ \lambda''_{113} & \lambda''_{213}  &\lambda''_{313}   \\\lambda''_{123}  & \lambda''_{223}  & \lambda''_{323} \end{array}\right)
=
\lambda''_{323} \left(\begin{array}{ccc} 3\times10^{-5}& 3\times10^{-3}&5\times10^{-2}\\
1\times10^{-4}&1\times10^{-2}&2\times10^{-1}\\6\times10^{-4}&5\times10^{-2}&1\end{array}\right).
\eeq
\item in the MFV framework \cite{Nikolidakis:2007fc,Csaki:2011ge}, the baryon number violating couplings depend just on $\tan\beta$ and an overall scale factor $w''$, while the lepton number violating operators are naturally suppressed. For $\tan\beta\gtrsim1$ we have:
\beq
&\lambda''_{ijk}=w'' \tan^2\beta\, m_i^{(u)}m_j^{(d)}m_k^{(d)}\epsilon_{jkl}V^*_{il}/v^3,\\\label{LMFV}
&\qquad\left(\begin{array}{ccc} \lambda''_{112}  & \lambda''_{212}  & \lambda''_{312}  \\ \lambda''_{113} & \lambda''_{213}  &\lambda''_{313}   \\\lambda''_{123}  & \lambda''_{223}  & \lambda''_{323} \end{array}\right)=
w''\tan^2\beta\left(\begin{array}{ccc}3\times10^{-12}&1\times10^{-8}&4\times10^{-5}\\6\times10^{-9}&1\times10^{-5}&6\times10^{-5}\\5\times10^{-7}&4\times10^{-5}&2\times10^{-4}
\end{array}\right).
\eeq
The coefficient $w''$ is not constrained by the flavor structure and should be an $\OO(1)$ number.
\end{itemize}
We take these examples as a justification to consider scenarios in which only Baryonic \rpv\ (BRPV) is allowed, while lepton number is conserved (at least to a good approximation). This is the scenario that will be studied in the rest of this paper. It should be noted that in both models (eqs. \eqref{LHOR} and \eqref{LMFV}), $\l''_{223}$ is the largest coupling that does not involve a top in the final state.

\begin{figure}
\begin{center}\includegraphics[width=5cm]{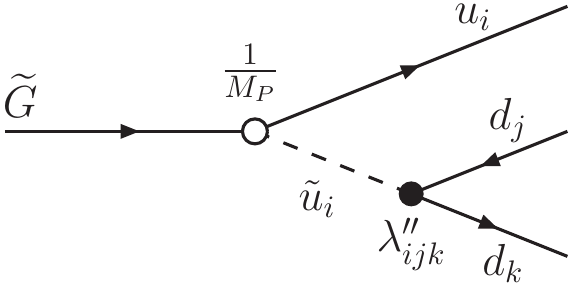}\end{center}
\caption{Gravitino RPV decay; the white vertex marks the $1/M_P$-suppressed interaction, while the RPV interaction $\lambda''_{ijk}\tilde{\bar  u}_i\bar d_j\bar d_k$ is marked by a black dot.}
\label{grvudd}
\end{figure}

Implicit in \rpa\ scenarios is stability of the lightest supersymmetric particle (LSP) which can provide a viable relic Dark Matter (DM) candidate. With a neutralino LSP, this is the usual SUSY WIMP scenario, and problems can arise from late time gravitino decays to the LSP \cite{Khlopov:1984pf}, disturbing the predictions of Big Bang Nucleosynthesis (BBN).  Alternatively, for a gravitino LSP, it is the NLSP decay to the gravitino that is suppressed by the Planck scale $M_P$ and can interfere with  BBN. In contrast, \rpv\ allows superpartners to decay directly and quickly into SM particles,\footnote{As noted above, we consider only baryonic RPV in this paper. Then, late NLSP decays are not a problem for a neutralino or squark NLSPs, but they can be for a stau NLSP. In the second case, heavy superapartners and/or light gravitinos would be needed.} solving this problem but at the same time eliminating dark matter candidates from the theory. If, however, the gravitino is the LSP, its decay (see Figure \ref{grvudd}) is suppressed by the SUSY breaking scale $F$ (or equivalently, by $M_P$),\footnote{For a comprehensive review of gravitino interactions, see Ref. \cite{Moroi:1995fs}.} by the \rpving couplings, and by the superpartner scale $\mt$.  This  naturally allows for lifetimes longer than the age of the universe \cite{Takayama:2000uz}. Because the gravitino is unstable, its decays will generate cosmic-rays and high-energy $\gamma$-ray emission which can potentially be detected by modern indirect-detection experiments. Given the non-observation of gravitino decay products, we will proceed to set limits on RPV couplings and will compare them to bounds coming from low-energy baryon-number-violating processes (which are especially weak for couplings involving third generation fields).
Although this has been studied in the literature, many groups have focused only on the bilinear RPV coupling $\mu_i L_i\phi_u$ \cite{Takayama:2000uz,Ishiwata:2008cu,Grefe:2014bta,Bobrovskyi:2010ps,Bertone:2007aw} with just Refs. \cite{Lola:2007rw,Lola:2008bk,Bomark:2009zm,Dal:2014nda}  discussing the trilinear interactions; weak scale supersymmetry was also frequently assumed. In this paper, we do not set the superpartner scale, we discuss the connection to models with flavor symmetries, which has been unexplored so far, and we show that the limits can be stronger than those from low-energy flavor physics.

Following Ref. \cite{Moreau:2001sr}, the decay rate of Figure \ref{grvudd} can be written as 
\beq\label{32gamma}
\Gamma_{3/2}\simeq \frac{19}{60\cdot 768\pi^3}\lambda''^2_{ijk}\frac{\mtr^3}{M_P^2}{\frud{\mtr}{\mt}}^4
\eeq
in the limit of vanishing masses for the final state particles and at leading order in $\mtr/\mt$. The lifetime is
\beq
\label{32lifetime}
\tau_{\tilde G\to u_id_jd_k}=2.9\times 10^{14} \text{sec} \left(\frac{\text{10 GeV}}{\mtr}\right)^3\frac1{\lambda''^2_{ijk}}\frud{\mt}{\mtr}^4.
\eeq
In this equation $\mt$ is the common mass scale of  the squarks which participate in the process; it is slightly  modified in presence of a large hierarchy between different squarks. In particular, the detailed dependence on the squark masses is recovered by substituting the factor $19\mtr^4/\mt^4$ in \eqref{32gamma} with
\beq
\frac{\mtr^2}{m_{\tilde u_i}^4}\left(3+2n_d\frac{m_{\tilde u_i}^2}{m_{\tilde d_j}^2}+3n_d^2\frac{m_{\tilde u_i}^4}{m_{\tilde d_j}^4} \right),
\eeq
where we have denoted by $\tilde d_j$ the lightest down squark, and $n_d$ is the number of down squarks participating in the process, $n=\system{1, \ m_{\tilde d_j}\ll m_{\tilde d_k}\\2,\ m_{\tilde d_j}\sim m_{\tilde d_k}}$.

With the pre-inflationary gravitino abundance washed out during inflation, gravitinos are produced  by thermal scattering at reheating and by decays of other fields (such as moduli, or the inflaton). We will show overclosure limits coming from the overproduction of gravitinos, and in the following we will assume $T_R>\mt>\mtr$ for the reheating temperature.
As a  conservative choice,  we will  assume that the full DM relic abundance is generated {\it in toto} at reheating;\footnote{If the universe reheats below $\mt$, gravitinos are not produced thermally. Still, a gravitino relic abundance might be produced by moduli or inflaton decay.}
 the second class of processes will just strengthen the overclosure bounds that we are considering.\footnote{On the other hand, these limits can be relaxed with a late entropy injection diluting the relic abundance.} The thermal scattering, with a cross section of order $\sigma\approx g_3^2 \frac{\mt^2}{M_P^2\mtr^2}$,  overcloses the universe unless (see \cite{Moroi:1995fs,Hall:2013uga} for the precise expression)
\beq\label{grvclose}
10^{-3}\frac{T_R\mt^2}{\mtr}\lesssim  M_PT_{eq}.
\eeq
where the equality holds if the gravitino forms all of the dark matter, as we will assume in the following, and $T_{eq}\sim 1.5$ eV is the temperature of matter-radiation equality.


\bigskip

This paper is organized as follows: in Section \ref{grvdecay}, we review the importance of \antid s for the indirect detection of dark matter candidates and the coalescence model of \antid \ formation.  We then compute and discuss the \antid \ injection spectrum.  In Section \ref{rpvlimits} we derive the upper limits on the RPV coupling $\l''_{223}$ from the lack of antideuterons and discuss the dependence on the SUSY and SUSY mediation scales. We apply these limits in Section \ref{flavor}, where we discuss the implications for models with flavor symmetries. Finally, we conclude in Section \ref{conclusions}.

\section{Antideuterons from Gravitino Decays}\label{grvdecay}

Measurements of the cosmic-ray antiproton spectrum by BESS~\cite{BESS,BESS2,BESS3} and PAMELA~\cite{PAMELA_ANTIPROTON} have provided important constraints on cosmic-ray transport in the galaxy, as well as placed limits on exotic source models such as dark matter annihilations or decays and primordial black hole emission. In the near future, data from the AMS-02 experiment on-board the international space-station will provide the most precise measurements to date.  While indirect detection limits on antiprotons currently provide the leading constraints on \rpving $\lambda''_{ijk}$ couplings, the production of secondary antiprotons through cosmic-ray spallation processes provides an astrophysical background with considerable uncertainty.

In 2000, Donato et al.~\cite{donato2000} proposed new physics searches, specifically neutralino annihilations, using heavier anti-nuclei such as antideuterons, antihelium-3, or antitritium.  In contrast to antiprotons, the secondary background for antideuterons is highly suppressed at low energies while gravitino decays produce a peaked spectrum. This happens for three reasons: first, the scattering of cosmic-ray protons with interstellar gas produces (secondary) antiprotons only if the cosmic-ray proton has a total energy above the production threshold $E_{\rm p}=7 m_p$ in the galactic rest frame.  At these energies, the density of Galactic cosmic-rays is substantial, and the antiproton spectrum below $\approx$~5~GeV becomes heavily populated by the astrophysical background.  In the case of antideuterons, this threshold is increased to $E_{\rm p}=17 m_p$, where a rapid decrease in the Galactic proton spectrum heavily suppresses the astrophysical background.\footnote{The proton spectrum peaks at approximately 10 GeV and subsequently falls off proportionally to $E^{-2.82}$.} Second, astrophysical production occurs in a center of mass frame which is highly boosted with respect to the rest of the galaxy, whereas dark matter decays occur at rest.  This results in the background spectrum peaking at higher energies than that of dark matter decays, which typically peaks in the non-relativistic regime. Finally, the small binding energy of antideuterons causes them to disintegrate rather than lose energy through inelastic scattering, unlike antiprotons for which such collisions lead to an increased abundance at lower energies.  These low-energy antideuteron astrophysical backgrounds are 10-50 times less in magnitude than the primary signals expected from naive thermal dark matter models~\cite{fornengo2013,ibarra2013}, so searches for low energy antideuterons can provide a promising discovery channel for new physics.  Heavier elements such as antihelium are even cleaner~\cite{duperray2005}, although the expected signals are too small to be observed with the current generation of experiments as shown by one of the authors (EC)~\cite{antihelium1} and later independently~\cite{antihelium2}.  For the Baryonic \rpving operators under consideration here, constraints on the couplings are currently competitive with those from antiprotons and are expected to improve substantially with the results of AMS-02.

The detection of antinuclei from dark matter decay has been thoroughly investigated in the literature with an emphasis on simple two-body final states such as $b\bar{b}$ or $W^+ W^-$ (see e.g. \cite{donato2000, profumo2005, cui2010,ibarraMC2013,kadastik,fornengo2013, ibarra2013} for antideuterons and \cite{antihelium1,antihelium2} for antihelium).  Recently, Ref.~\cite{Dal:2014nda} provided the first \antid\ constraints for gravitinos decaying through a variety of \rpving operators.  One novel feature of their analysis is the detailed treatment of the Monte Carlo parameters controlling the hadronization model which are tuned to reproduce a wider array of experimental antideuteron production rates.  In this paper, much of the same production and propagation framework is used, but we do not vary the hadronization model in order to extract the model-\emph{dependent} features of gravitino decay, and compare them to standard treatments of decaying dark matter.  In doing so, we can provide simple scaling relations which allow BRPV coupling constraints to be easily adapted from future updated measurements and more sophisticated propagation schemes that are presented in the context of two-body decays to heavy quark pairs. 

In any process producing antinucleons, it is possible for antiprotons and antineutrons to bind together into a nucleus and produce antideuterons.  The traditional formation model, known as the `coalescence mechanism', was designed to empirically describe nuclei production in heavy-ion collisions based on the phase-space distributions of the constituent nucleons.  It possesses a single energy-independent parameter, the coalescence momenta $p_0$, and assumes that if any antineutron and antiproton pair have relative invariant 4-momenta $(k_n-k_p)^2=(\Delta \vec{k})^2 - (\Delta E)^2 \leq p_0^2$, they will fuse and form an antideuteron.  The parameter $p_0$ is then tuned to match collider measurements of \dbar \ production.

It has long been known that this model cannot accommodate the available data for a single value of the coalescence momenta to better than a factor of $\sim 3$.  Despite this simplistic model, an improved prescription is largely hindered by limited collider data for production of antideuterons from $e^+e^-$ collisions at high energies, as well as a lacking understanding of the underlying nuclear formation dynamics.  However, recently renewed interest in antideuteron searches have led to at least two important improvements.  First, it was pointed out in Ref.~\cite{kadastik} that the isotropic nucleon distribution functions used in analytic estimates of formation rates led to an artificial suppression of the \dbar\ production rate at large center of mass energies.  In particular, the jet structure of high-energy showers introduces significant angular correlations between nucleons.  One must therefore run Monte Carlo simulations and apply the coalescence mechanism on an event-by-event basis using the simulated phase space distributions of protons and neutrons.  Second, it was realized that the antideuteron wave-function is spatially localized to $\approx 2$~fm and contributions to the nucleon population from long-lived baryons should be omitted, as they decay at large relative distances from the other particles in the shower.  In practice, weakly decaying baryons are then excluded by stabilizing particles with a lifetime $\tau>$~2fm/c with a negligible dependence on this parameter due to the large gap between weak and hadronic timescales. 

For our study, we first use {\tt Feynrules v2.0} package \cite{Alloul:2013bka} (using a modified version of the \texttt{gld-grv} \cite{Christensen:2013aua} and \texttt{RPV-MSSM} \cite{Fuks:2012im} model files)  to translate our \rpving Lagrangian into a UFO format readable by matrix element generators.  The matrix elements and phase space for the hard process $\tilde{G}\to\bar{u_i} \bar{d_j} \bar{d_k}$ are then generated using {\tt MadGraph v5.0} and {\tt MadEvent}~\cite{Alwall:2011uj}.  Finally, these parton level distributions are fed into {\tt Pythia 8.1}~\cite{Sjostrand:2007gs} for showering and hadronization.

In order to fix the coalescence parameter we must choose a value which reproduces a measured rate. As previous studies have noted, a single value of the coalescence momentum cannot simultaneously reproduce rates from different underlying processes such as pp vs $e^+e^-$.  While this can be slightly improved by tuning the hadronization parameters, we follow previous studies which use electron-positron collisions more likely to resemble a dark matter scenario -- i.e. color singlets that are not composite.  Following the approach of Refs. \cite{ibarra2013,fornengo2013,cui2010}, we use $e^+e^- \to$~\dbar\ measurements from ALEPH at the $Z^0$ resonance, finding $(5.9\pm 1.8 \pm .5) \times 10^{-6}$ antideuterons per hadronic $Z^0$-decay with \dbar\ momenta 0.62-1.03 GeV/c and polar angle $|\cos{\theta}|<0.95$ (\cite{aleph}).  We find a value $p_0^{A=2}=0.192\pm .030$ GeV/c consistent with Refs. \cite{ibarra2013,fornengo2013}.

In Figure \ref{fig:injection} we show the typical antideuteron injection spectra for a gravitino decay of mass $m_{3/2}=$10 GeV, 30 GeV, 100 GeV, 1 TeV, and 10 TeV.  In solid lines, we show the spectra from the heaviest accessible channel, which is expected to dominate the decay rate in scenarios with flavor symmetries, while dashed lines show the second heaviest contribution\footnote{In the case of the $cbs$ channel at 10 GeV we observe no events.}. For comparison, we also show the spectra for a standard dark-matter decay to $b\bar{b}$ in dotted lines.  Shaded bands show the acceptance energies for BESS (red), GAPS (green), the low-energy band of AMS-02 (blue), and the high energy band of AMS-02 (gray).  Here we assume that the spectra will be shifted to lower energies as the antideuterons propagate through the heliosphere and shift each band upward in energy due to the Fisk potential $\phi_f=500$~MV acting on a unit electric charge in accordance with the Gleeson \& Axford Force Field approximation~\cite{forcefield}.  The vertical normalization of each energy band is arbitrary and we have slightly offset the BESS band in order to keep the others visible.  We note that while the energy range of each experiment is fixed, they are rescaled by a factor $m_{3/2}^{-1}$ in these dimensionless coordinates. With the injection spectra now in hand, several observations can be made:

\begin{figure}[t]
\begin{center}
\includegraphics[width=\textwidth]{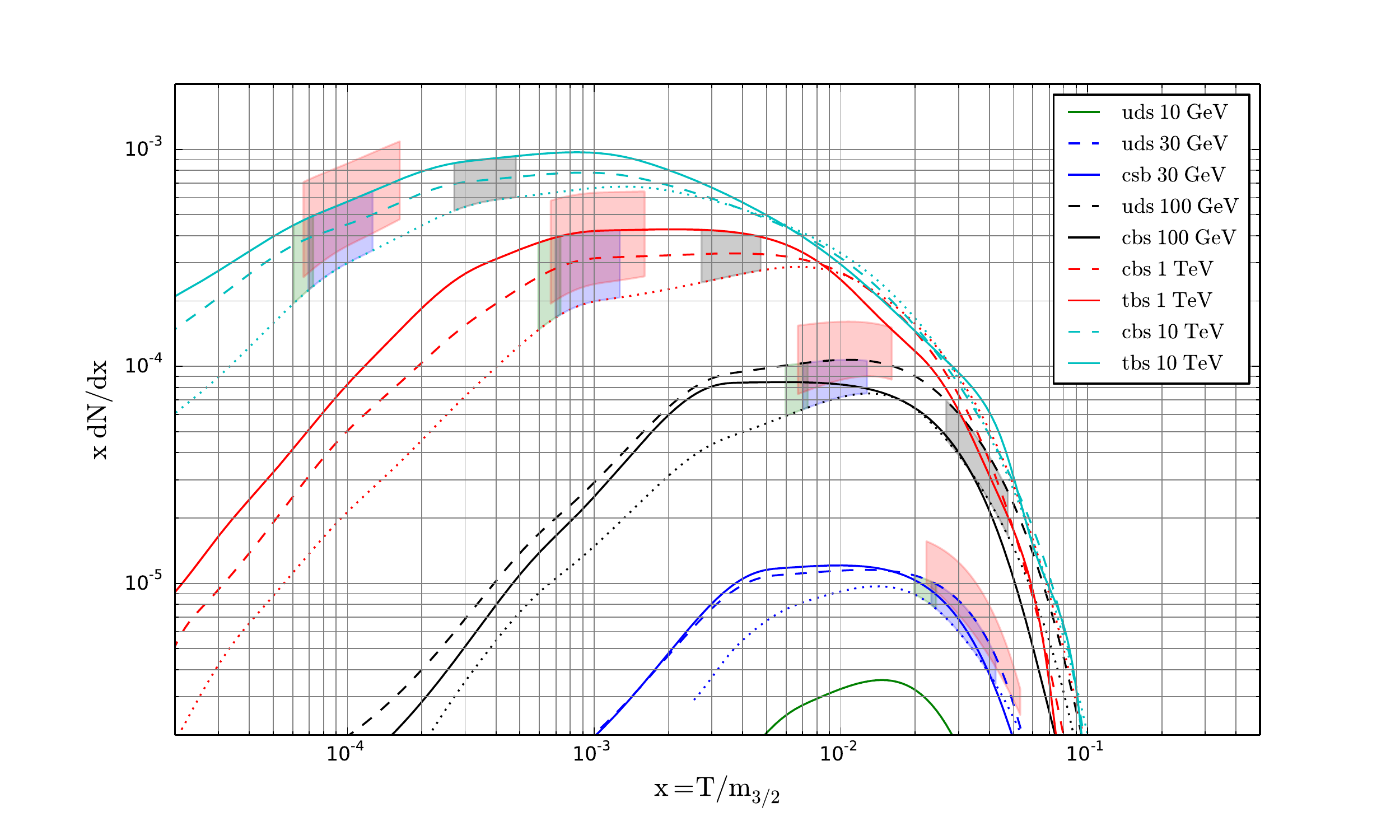}
\end{center}
\caption{Antideuteron injection spectra for  different operators involved and different gravitino masses, as  displayed in the legend. $dN$ is the average number of antideuterons with energy $dT$ generated in the decay of a single gravitino.
In particular, we display spectra generated by the operators $\bar u_1\bar d_1\bar d_2$ ($uds$), $\bar u_2\bar d_2\bar d_3$ ($cbs$), $\bar u_3\bar d_2\bar d_3$ ($tbs$).  Solid lines represent the heaviest accessible channel while dashed lines show the second heaviest.  Dotted lines represent the case of a 2-body decay to $b$-quarks, which is often presented in antideuteron analyses.  In shaded bands, we show the ranges of experimental detectability after accounting for solar modulation effects.  The bands are for BESS (red), GAPS (green), AMS-02-L (blue), and AMS-02-H (gray).  We have vertically offset the BESS band for readability (vertical normalizations for these bands are arbitrary). The energy range of the bands is identical for different gravitino masses, but in these coordinates the horizontal locations scale as a function of $m_{3/2}^{-1}$.}
\label{fig:injection}
\end{figure}

\begin{enumerate}
\item Comparing between decay channels, we see that the second lightest quark mass channels have a significantly harder spectrum than the heaviest. For $m_{3/2}$ less than a few hundred GeV, these low mass final state channels yield slightly more detectable antideuterons.  Such behavior is also evidenced in Ref.~\cite{Dal:2014nda} where the light quark channels provide the best limits on the trilinear BRPV coupling. Interestingly, this behavior reverses for $m_{3/2}\gtrsim$\ 1 TeV, where the heaviest quark channel dominates by a factor $\sim 20-30\%$ over the detectable low energies. One explanation may be the following: Increased jet multiplicity as the 2nd and 3rd generation quarks cascade down to $u$ and $d$ type quarks will divide the gravitino's energy.  For low masses, this could sufficiently raise the threshold where heavy channels can consistently form the requisite number of protons and neutrons. When the gravitino mass is very high, each jet will contain energy $E\gg m_p$, and the 3-tiered decay of the top-quark will effectively soften the otherwise harder spectrum. 

\item Compared with the 2-body decay to $b\bar{b}$, we see a significantly softer spectrum for our gravitino decay in all cases.  This results in a mild enhancement in detectable antideuterons of $\mathcal{O}(50\%)$ for $m_{3/2}\approx 50$ GeV increasing to a significant factor $\approx$ \ 3 above 1 TeV.  This is mostly attributed to the higher initial multiplicity of quarks in the final state of the hard process which splits the initial gravitino energy into three final states rather than two.  In addition to this, the 3-body phase-space allows the hard jets to occasionally align, and thus increase the probability of a neutron and proton coalescing.  In the 2-body case, jets are forced back-to-back for a decay at rest, and are therefore less likely to have cross-jet correlations. 

\item The formation model used here is distinct from Ref.~\cite{Dal:2014nda}. Notably, we use {\tt Pythia} for hadronization (based on the string fragmentation model) while in Ref.~\cite{Dal:2014nda}, {\tt Herwig++} (based on the cluster hadronization model) is used. It has been shown in Ref.~\cite{Dal:2012my} that differences between the two different models can lead to substantially different preferred values of the coalescence momentum and variances in the spectrum of anti-deuterons produced. Furthermore, our coalescence momentum is fit to a single data-point at the $Z^0$-resonance while the Ref.~\cite{Dal:2014nda} varies the parameters of the hadronization model in order to reproduce results from $e^+ e^-$ and $p p$ collisions at 50 GeV-7 TeV.  We therefore expect to see some level of disagreement at higher gravitino masses. In fact, we do find a significant enhancement in our yield (integrated over the low-energy experimental bands) of around 30\% at 50 GeV up to 300\% at 1 TeV.  As this is an artifact of the underlying hadronization and coalescence model, it occurs independent of the two results enumerated above for which the comparison is based on a common framework.

\end{enumerate}

In order to translate the injection spectra into the observable astrophysical fluxes, one must propagate the \dbar\ nuclei through two stages: interstellar transport from the position of production to the solar system and modulation of the spectra during through the heliosphere.  Interstellar transport of antiprotons and light-nuclei is very well studied but unfortunately still suffers from considerable uncertainties.  Our implementation of interstellar propagation follows the standard semi-analytical treatment using the `two-zone diffusion model' which provides a simplified but good approximation by neglecting energy losses and diffusive reacceleration.  The neglect of tertiary processes -- i.e. non-annihilating inelastic scatters are treated as annihilations -- is well justified and in particular does not redistribute the spectrum of antideuterons toward lower energies.  In other words, the Green's function which solves the simplified transport equation is proportional to a $\delta$-function and as a result, the injection spectrum can be factored out of propagation.  Similarly, propagation through the solar system in the Force-Field model \cite{forcefield} only introduces an energy dependent scaling and a global shift of the spectrum to $500$~MeV lower energy.  This implies that event-rates and observable fluxes can be readily compared by computing the ratio of the injection spectra, integrated over the experimental acceptance range.  As AMS-02 and GAPS results become available, propagation uncertainties are likely to be reduced based on upper limits on the antiproton spectrum.  Recent analyses have already incorporated sophisticated numerical treatments of interstellar and heliospheric propagation and present their findings in terms of annihilation or decay to heavy quarks~\cite{fornengo2013}.  Conversion of these rates to case of gravitino dark matter is therefore a simple rescaling according to the integrated ratios of injection read from Figure~\ref{fig:injection}.  Our treatment of propagation and conversion of the flux to event rates is completely identical to that presented in Section 3 of Ref.~\cite{Dal:2014nda} and we therefore omit our own details, pointing the interested reader to the discussion presented there\footnote{Here, and in Ref.~\cite{Dal:2014nda}, the halo model chosen is a standard NFW profile and the Fisk potential is taken to be $\phi_F=500MV$.  In the next section, our results use the `MED' propagation model to compute the flux, although it should be kept in mind that these propagation uncertainties span 2-3 orders of magnitude. See also Ref.~\cite{ibarra_review} for a recent review of indirect detection of decaying dark matter.}.  In the next section we will discuss how the differences between our injection spectra and that of Ref.~\cite{Dal:2014nda} lead to different limits on the BRPV couplings.

\section{Model-independent limits on RPV couplings}\label{rpvlimits}
In this section, we will show the model-independent limits on the RPV couplings coming from null observation of antideuterons at the BESS experiment \cite{Fuke:2005it} and the future reach of the AMS-02 experiment; as a starting point, we take the limits on $\l''$ due to lack of observation of cosmic antideuterons from Ref.~\cite{Dal:2014nda}, where the full gravitino decay $\tilde G\to \bar u_i \bar d_j \bar d_k \to \bar D+\ldots$ was studied. There, no flavor symmetry was assumed, and the strongest limits were set on the coupling $\l''_{112}$ ($uds$ channel) in the range $10 \gev\leq \mtr \lesssim 1\tev$, for $\mt=1\tev$.

In Figure \ref{fig:injection}, we showed the spectra generated by decays of a gravitino in the $uds$, $cbs$, and $tbs$ channels for different choices of the gravitino mass: compared to Ref. \cite{Dal:2014nda}, we find that the number of \antid s produced in the \uds\ channel in the BESS energy sensitivity range is about  50\% higher at low gravitino masses ($\mtr=50\gev$), and about a factor of 3 higher at $\mtr=800\gev$.
The local flux of antideuterons scales as $\l''^2$.  A change in the injection spectra by a factor $A$ therefore strengthens the bounds on $\l''$ by a factor of $\sqrt A$ assuming we are in the signal dominated regime (i.e. low relative background flux). We then rescale the 95\% confidence level limits on the coupling $\l''_{112}$ from Figures 7 and 8 of \cite{Dal:2014nda} in order to self-consistently employ our injection spectra.

As previously mentioned the number of antideuterons produced in the \uds\ and \cbs\ channels differs only by 20-30\% in the energy range accessible by the BESS and AMS-02 experiments. This implies that the limits on the respective couplings differ only by 10-15\%. 
We  are particularly interested in the coupling $\l''_{223}$, which, according to both eqs. \eqref{LHOR} and \eqref{LMFV}, is the largest coupling that does not involve a top in the final state. As such, the decay process will proceed through the $cbs$ channel for gravitino masses between the $b$-quark mass and about 1\tev.  At high gravitino masses (above 1 TeV), the decay involving the coupling $\l''_{323}$ and a top quark will also be relevant. For such high masses, the \tbs\ channel gives a slightly higher number of antideuterons when compared to the \cbs\ channel, so that the 95\% CL limit on $\l''_{323}$ will be slightly stronger than the limit on $\l''_{223}$ at the same scale.  As the resulting antideuteron spectrum flattens at the low energies relevant experimentally, we can extend the bounds from Ref. \cite{Dal:2014nda} to gravitino masses above 1\tev\ and expect no qualitative change in behavior. As we will see in Section \ref{flavor}, because both flavor models predict $\l''_{323}>\l''_{223}$, $\l''_{323}$ will be the most constrained in this regime. At lower masses, the constraints are strongest for $\l''_{223}$.

\begin{figure}[tb]
\begin{center}
\includegraphics[width=8.5cm]{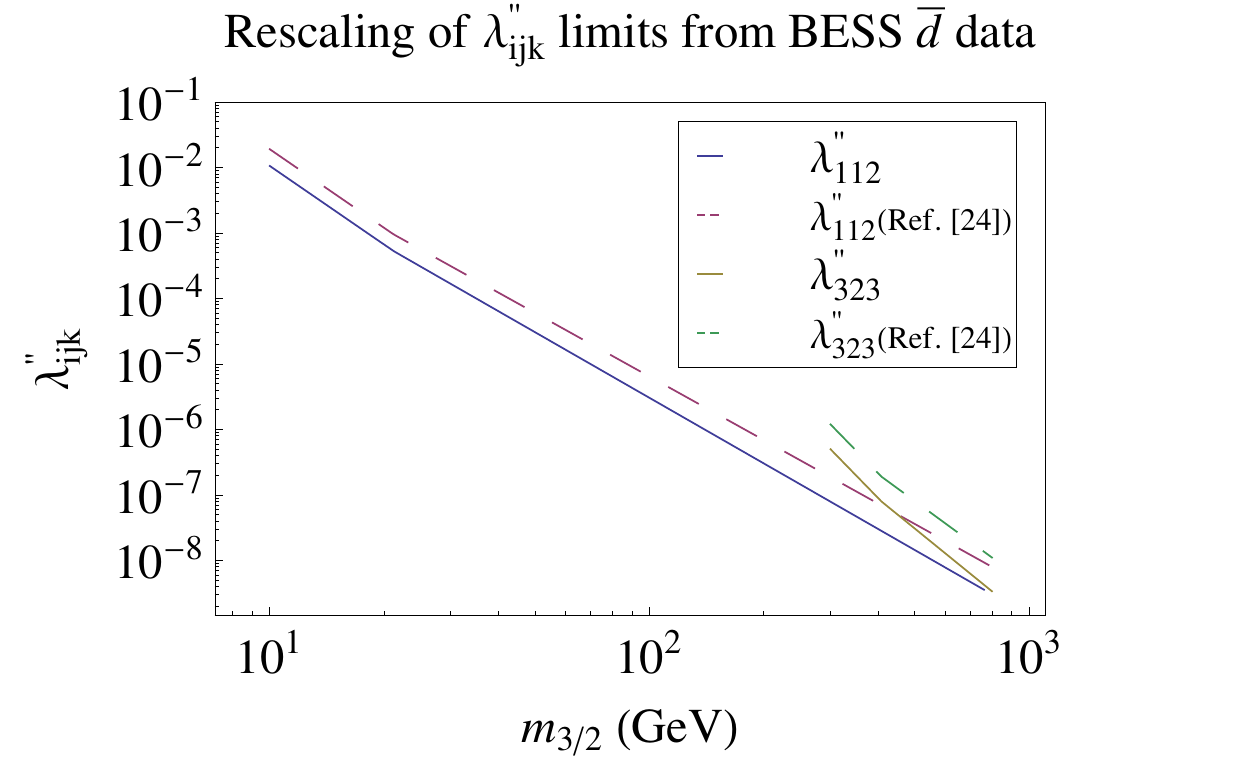}
\caption{95\% CL upper limits on different couplings $\lijk$ from antideuteron data, compared to the values in Ref. \cite{Dal:2014nda}. The region above the lines is excluded. In this graph, $\mt$ is set at 1\tev.
}
\label{fig:singlelambda}
\end{center}
\end{figure}

In Figure \ref{fig:singlelambda} we compare the limits from Ref. \cite{Dal:2014nda} with the ones that will be used in the following. We plot the bounds on the individual couplings $\l''_{112},\l''_{323}$ ($\l''_{223}$ is degenerate with $\l''_{112}$) as a function of the gravitino mass, with a reference superpartner scale $\mt=1\tev$. 

\begin{figure}[tb]
\begin{center}
\includegraphics[width=13cm]{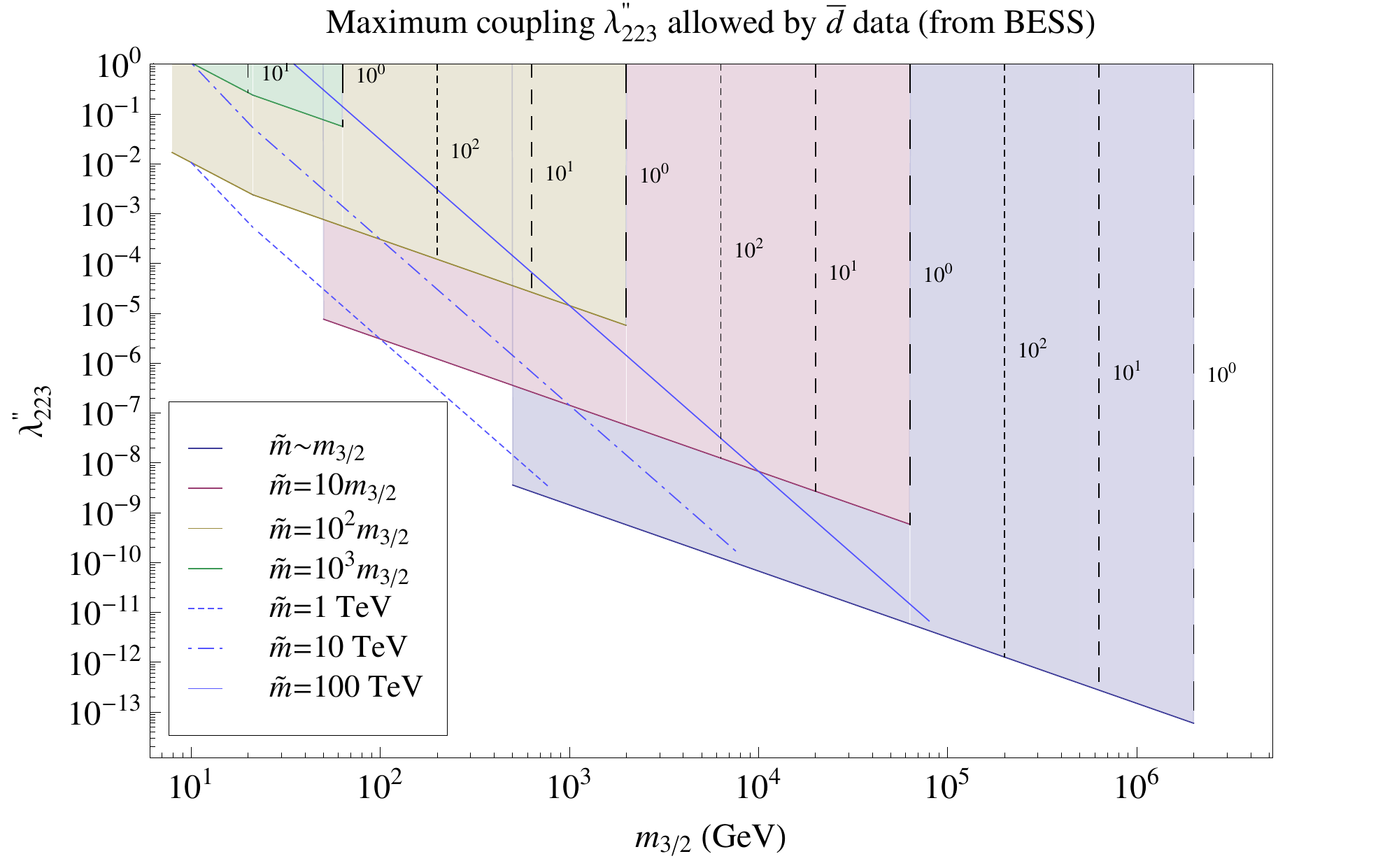}
\caption{Maximum coupling $\lambda''_{223}$ allowed by the non-observation of antideuterons at BESS, for different values of $\mtr/\mt$; the shaded region above each continuous line is excluded. The  blue lines  correspond to fixed values of $\mt=1,\,10,\,100$ TeV (repectively dotted, dashed and solid  line). The vertical dashed lines are the upper  limits  on the gravitino mass coming from overproduction of gravitinos at reheating, with the labels indicating the respective values of $T_R/\mt$; the regions to the right of these lines are excluded for each given value of $\mtr/\mt$.
}
\label{LmaxTR}
\end{center}
\end{figure}

Provided that the gravitino is the LSP, the injection spectra are independent of the mass of the superpartner  involved in the process. Thus, the only dependence on the superpartner scale $\mt$ is in the hard process that determines the decay rate, as shown in Eq.~\eqref{32gamma}. In Figure \ref{LmaxTR}, we show how the limits on $\l''_{223}$ depend on the superpartner scale: the dot-dashed diagonal lines show the upper limits on $\l''_{223}$ for given values of $\mt=1,10,100$ TeV. The parameter space above each line is ruled out. Alternatively, we fix the ratio $\mt/\mtr$ to approximately $1,10,10^2,10^3$ and show the allowed parameter space. Here the vertical black dashed lines show the upper bounds on the gravitino mass coming from overproduction during reheating. 
Setting the ratio $\mt/\mtr$ corresponds to setting the SUSY mediation scale $M$: if the gravitino mass is $\mtr\approx \frac F{M_P}$ and the squark masses are $\mt\sim \frac F M$, we have $\mt/\mtr=M_P/M$. The limits on $\l''_{223}$ presented in Figure \ref{LmaxTR} are independent of the flavor structure. As stressed above, for $\mtr\gtrsim 1\tev$, similar limits apply to $\l''_{323}$.

It should be noted that the relevant squarks in this process are  $\tilde s_R,\tilde c_R, \tilde b_R$; limits from  \rpa\ conserving LHC searches for first and second generation squarks are above 1 TeV, while for $\tilde b_1$ they are at $\approx 650$ GeV \cite{Aad:2013ija,Chatrchyan:2013lya}; without \rpa\ it is {\it in principle} possible for squarks to be significantly lighter than the \rpa\ conserving constraints. However, we find it  a plausible assumption that the superpartners are not hiding at extremely low masses. Allowing for a little hierarchy between $\tilde b$ and $\tilde c,\tilde s$, we require that $\mt\geq500$ GeV for simplicity.

\section{Constraints on models with flavor symmetries}\label{flavor}

As  seen in eqs. \eqref{LHOR} and \eqref{LMFV}, flavor symmetries constrain the structure of the unknown RPV couplings; limits on one coupling (as $\l''_{223}$) directly translate into limits on all the other couplings (in particular, the largest one, $\l''_{323}$).

\subsection{Horizontal Symmetries}
In models with a horizontal symmetry, as $\l''_{223}$ is smaller than $\l''_{323}$ by a factor of 20, the conversion is straightforward: for $\mtr$ between $10$ GeV and 200~GeV, the \cbs\ channel is predominant in creating antideuterons. Above the top mass, from about 200\gev\ to about 1 TeV, the \tbs\ channel contribution grows until it eventually outweighs the \cbs\ channel due to its larger coupling. For $\mtr\gtrsim 1 \tev$, the \tbs\ channel  gives approximately 20-30\% more antideuterons per decaying gravitino than the \cbs\ channel  (see Figure \ref{fig:injection}), with \tbs\ dominating the decays given the larger coupling.

\begin{figure}[t]
\begin{center}
\includegraphics[width=7.5cm]{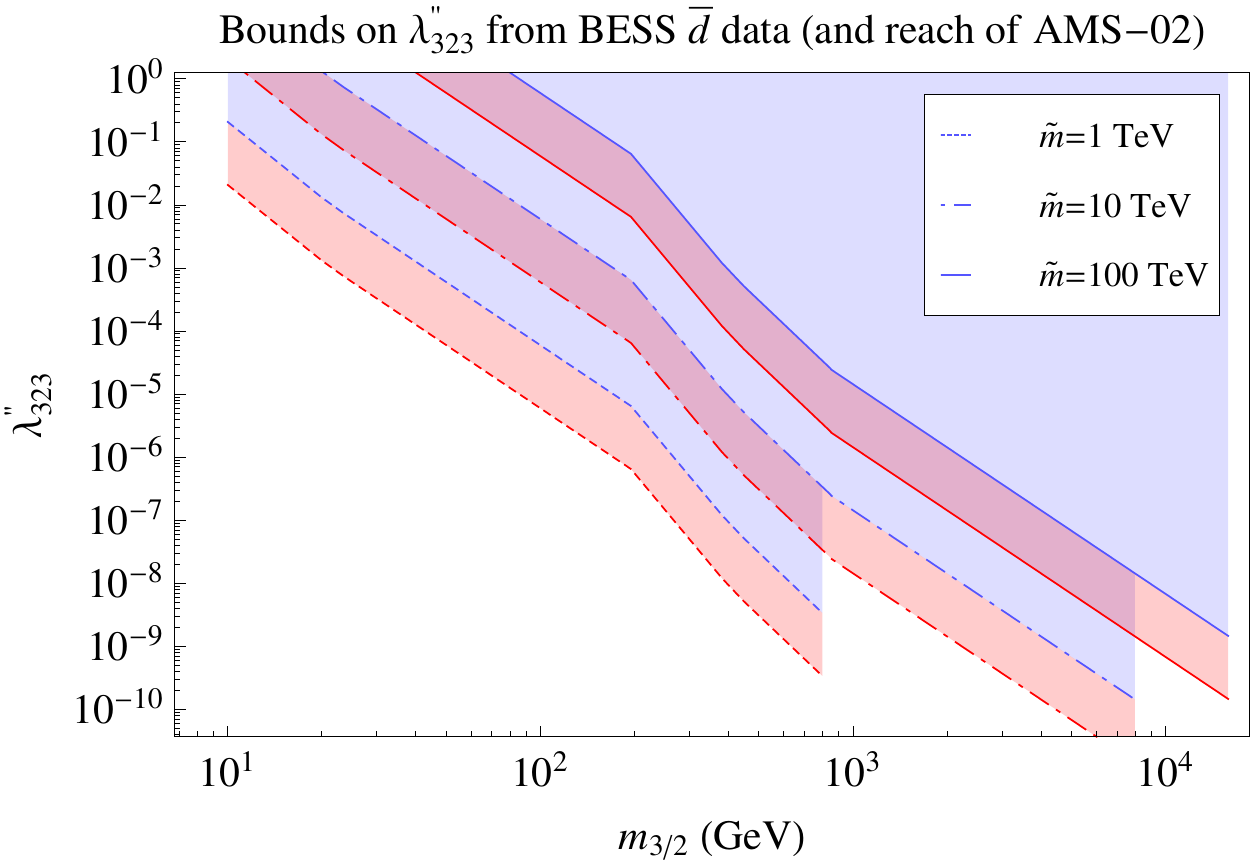}
\includegraphics[width=7.5cm]{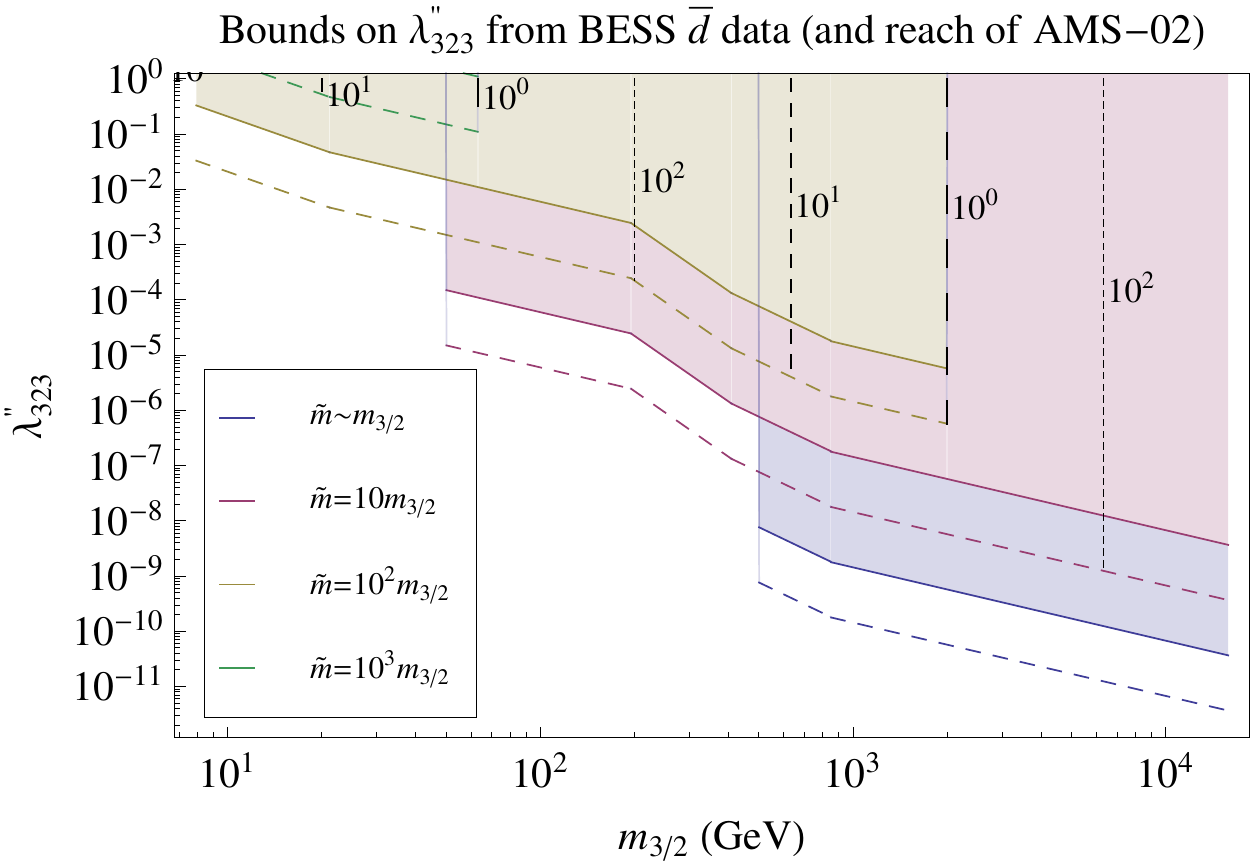}
\caption{Maximum coupling $\lambda''_{323}$ allowed by the non-observation of antideuterons at BESS, and the reach of AMS-02, in the context of horizontal symmetries. The region above each line is excluded. Left: the blue  lines indicate current upper bounds on $\lambda''_{323}$ for fixed values of $\mt=1,\,10,\,100$ TeV, while the red lines and red shaded area correspond to the parameter space which will be probed by AMS-02. Right: for different values of $\mtr/\mt$, solid lines show the upper bound on $\lambda''_{323}$ from BESS, while dashed lines show the future reach of AMS-02. The vertical dashed lines are the upper  limits  on the gravitino mass coming from overproduction at reheating, with the labels indicating the respective values of $T_R/\mt$; the regions to the right of these lines are excluded for each given value of $\mtr/\mt$.}
\label{Lhoriz}
\end{center}
\end{figure}

In Figure \ref{Lhoriz}, we present the limits on the largest allowed RPV coupling, $\l''_{323}$, which is likely to be the most relevant for LHC phenomenology. On the left, we show limits on $\l''_{323}$ for given values of $\mt$, while on the right we fix the ratio $\mtr/\mt$. We also show the future reach of the AMS-02 experiment. An improvement of a factor of 10 is expected across the entire range of gravitino masses.

These limits on RPV couplings can be compared to those found when requiring RPV gives a small contribution to low-energy flavor changing processes, in particular neutron-antineutron oscillation or the neutron decay $n\to \Xi$. In \cite{Monteux:2013mna}, one of the authors (AM) showed that the largest RPV coupling is bound to be less than about $10^{-2}-10^{-3}$, depending on the dominant process and the superpartner scale, and independent of the gravitino mass. We see that, apart from $\mtr\lesssim30-50\gev$,  the antideuteron limits from a decaying gravitino DM {\it are stronger than those from low-energy experiments}.

Some reference scales should be kept in mind while discussing these limits:
\begin{itemize}
\item $\l''_{323}=10^{-7}$; in Ref.~\cite{Barry:2013nva}, it was discussed how large \rpv\ would have washed out baryon number in the early universe if the $B$-violating processes were in equilibrium at a temperature of order $\mt$, and how RPV SUSY at colliders would most likely involve displaced vertices (this was also pointed out in \cite{Monteux:2013mna} in the context of horizontal symmetries, and in \cite{Krnjaic:2013eta}). In Figure \ref{Lhoriz}, the requirement $\l''_{323}<10^{-7}$ is automatically satisfied for $\mtr\gtrsim 500 \gev$ for TeV-scale SUSY. For heavier superpartners, or split spectra, it is true for $\mtr\gtrsim 2-5$ TeV. In other words, for large splittings between the gravitino and the superpartners, the cosmic ray flux from gravitino DM is more constraining than the requirements of having baryogenesis with a large reheating temperature. It should be noted that in baryogenesis scenarios with low reheating temperature \cite{Affleck:1984fy,Dimopoulos:1987rk}, this bound does not apply as the baryon asymmetry is created after the BNV processes has fallen out of equilibrium (An alternative setting in which baryogenesis is generated by the decay of a meta-stable WIMP was presented in \cite{Cui:2012jh}.). Given the BICEP2 detection of a tensor-to-scalar ratio $r=0.2$ \cite{Ade:2014xna}, it remains to be seen if such scenarios are still viable.

\item  $\l''_{323}=10^{-9}$: in \cite{Monteux:2013mna} one of the authors (AM) showed that, in order to evade collider signatures for subTeV SUSY, the lower limit $\l''_{323}>10^{-9}$ should hold for either a neutralino NLSP (in which case the missing energy signature of \rpa\ conserving SUSY  reappears) or a stop NLSP (for which the the long lived stop  hadronizes into R-hadrons and heavy stable charged particle searches would apply). From Figure \ref{Lhoriz} we can conclude that heavy gravitinos with $\mtr\sim\mt>1\tev$ imply $\l''_{323}\lesssim10^{-9}$ and either give standard \rpcing LHC phenomenology or long-lived particles.

\item $\l''_{323}=10^{-13}$: the lowest scale for which the RPV decay of the NLSP happens before BBN is $10^{-13}$. We see that this scale is not particularly constrained by Figure \ref{Lhoriz}.
\end{itemize}

For collider-accessible superpartners, we can conclude that if the coupling $\l''_{323}$ was measured to be large, it would imply a small gravitino mass.

\subsection{MFV: A gravitino on the edge}
In models with a minimal flavor violating structure \cite{Nikolidakis:2007fc,Csaki:2011ge}, the only free parameters are the overall scale $w''$, $\tb$ and $\mtr$. 
The \cbs\ coupling $\l''_{223}$ is set to
\beq
\l''_{223}=4\times 10^{-5}\tan^2\beta w''
\eeq
As we will see, we do not need to consider the larger \tbs\ coefficient $\l''_{323}\simeq5\l''_{223}$ as the limits from the \cbs\ channel are already enough to rule out the large gravitino mass range where the \tbs\ channel  (with a top quark in the final state) would dominate.

\begin{figure}[t]
\begin{center}
\includegraphics[width=7.5cm]{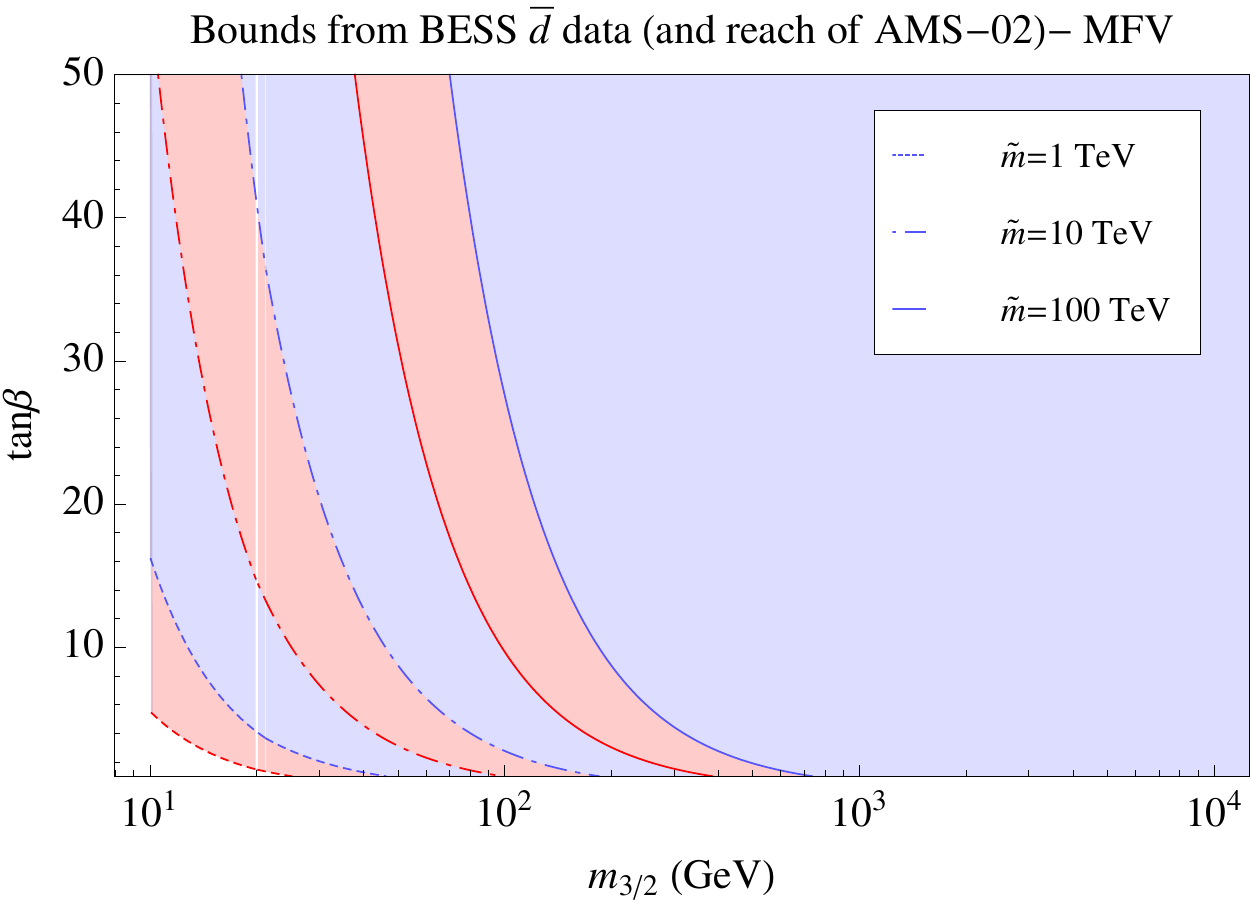}
\includegraphics[width=7.5cm]{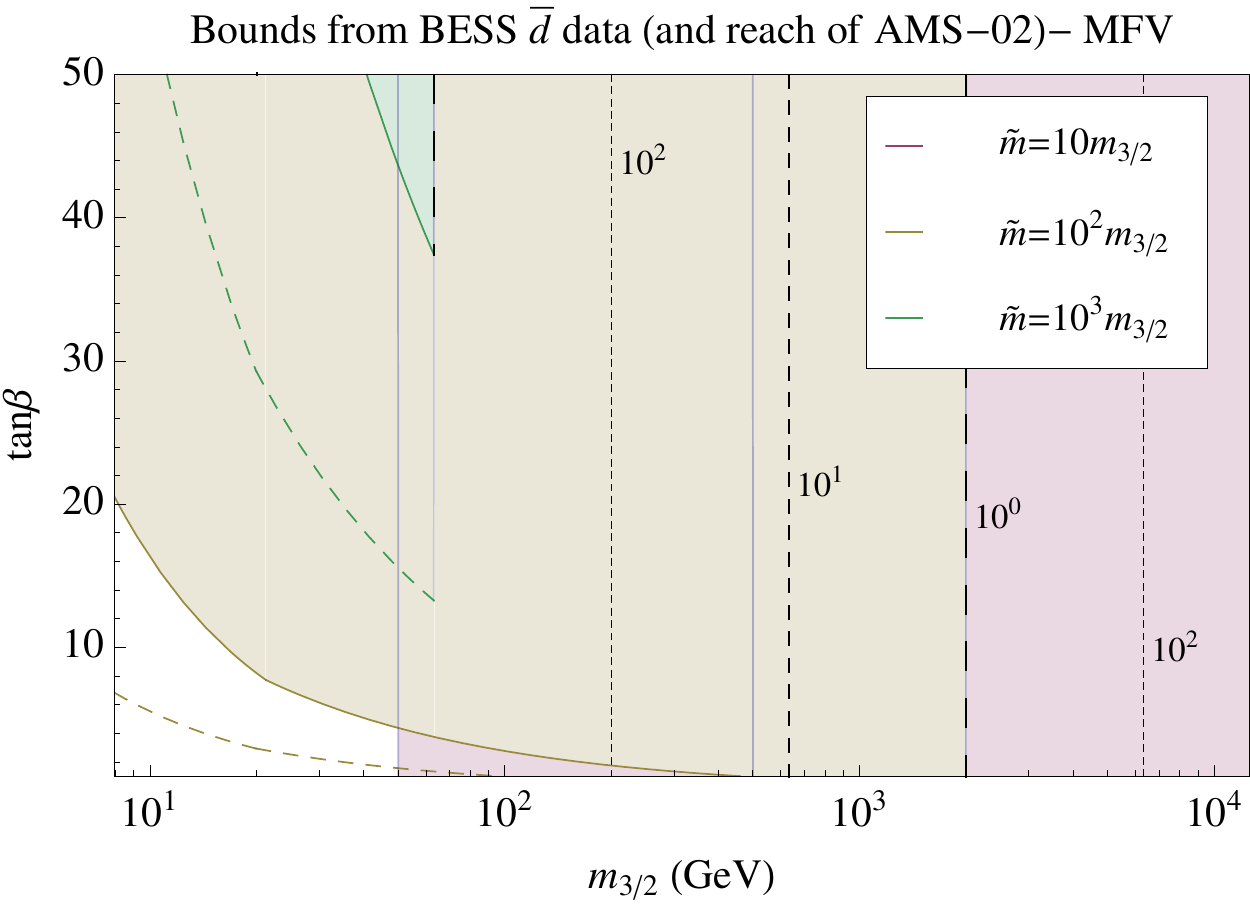}
\caption{Constraints in the $\mtr-\tan\beta$ plane when the RPV couplings have a MFV flavor structure; the shaded region above each continuous line is excluded in that case. The region above each line is excluded.. Left: the blue lines indicate current upper bounds on $\lambda''_{323}$ for fixed values of $\mt=1,\,10,\,100$ TeV, while the red lines and red shaded area corresponds to the parameter space which will be probed by AMS-02. Right: for different values of $\mtr/\mt$, solid lines show the upper bound on $\lambda''_{323}$ from BESS, while dashed lines show the future reach of AMS-02. The vertical dashed lines are the upper  limits  on the gravitino mass coming from overproduction at reheating, with the labels indicating the respective values of $\frac{T_R}{\mt}$. We set $w''\sim1$; the limits scale as $1/\sqrt{w''}$.
}
\label{MFVm32tb}
\end{center}
\end{figure}

Setting $w''\sim1$, the resulting limits on $\mtr$ and $\tb$ are shown in Figure \ref{MFVm32tb}. For fixed values of  $\mt$, we are forced into a corner with small $\tb$ and/or small $\mtr$. In particular, for LHC-accessible superpartners, the gravitino must be lighter than 50\gev\ if $\tb\sim1$, and $\tb$ can be as large as 20 for $\mtr =10$ GeV. We also note that it is possible to accommodate a 125 GeV Higgs mass in the case of large $\tan\beta$ and $\mt=10$ TeV, as well as in the case of $\mt=100$ TeV. The AMS-02 experiment will remove a large fraction of this parameter space: for $\mt=1\tev$ the limits will be $\mtr\lesssim10\gev$, with $\tan\beta\lesssim5$ for $\mt\sim10\gev$. 

For given values of $\mtr/\mt$, the only viable options are $\mt=10^2\mtr$ and $\mt=10^3\mtr$. In the first case, the gravitino mass should be lower than $\sim 200\gev$ for low $\tb$ and below a few tens \gev\ for larger $\tb$; in the second case, a larger zone of the parameter space will be explored by AMS-02, but the gravitino is easily overproduced.

If the overall scale factor $w''$ was allowed to be $\ll1$ a larger region of the parameter space would survive. Given that some couplings are larger than $\OO(10^{-7})$, the MFV structure is consistent with high temperature baryogenesis only if $w''\ll1$. In this case, the limits would scale as $\sqrt{w''}$ and can be relaxed.  Still, $w''$ cannot be infinitely small, and using the expression \eqref{LMFV} for the RPV couplings, we avoid the previously discussed limit of $\l''_{323}\gtrsim 10^{-9}$ with $w''\gtrsim10^{-5}$. If allowed, a small $w''$  should be considered as a tuning of the model.

\section{Conclusions}\label{conclusions}

In this work, we studied how the non-observation of antideuterons cosmic rays places significant constraints on gravitino dark matter in baryonic \rpving models with flavor symmetries. We studied a selected number of decay channels and presented limits on the RPV couplings $\l''_{223}$ and $\l''_{323}$, which are almost everywhere stronger than bounds from baryon-number-violating low-energy processes. If flavor symmetries can be used as guides, these are the largest couplings and severe bounds can be cast. While the limits on horizontal flavor symmetries are not as strong, in the minimal flavor violating case the gravitino mass is forced to $\mtr\lesssim 20\gev$ for TeV-scale SUSY. The AMS-02 experiment will be able to reduce this bound below 10\gev. A suggestive implication (which could hold at least for the MFV scenario) is that the gravitino might be effectively stable, not because of a discrete symmetry such as \rpa, but because decays are not kinematically allowed.\footnote{If the gravitino is lighter than the proton, the proton can decay to it $p\to K^+ \tilde G$. This was considered in Refs.~\cite{Choi:1996nk,Choi:1998ak}), with the most relevant bounds being:
\beq
\l''_{112}\leq 5\times 10^{-16}\fru \mt{300\gev}^2\fru{\mtr}{1\text{ eV}},\qquad
\l''_{323}\leq 5\times 10^{-8} \fru \mt{300\gev}^2\fru{\mtr}{1\text{ eV}}
\eeq
For models with a horizontal symmetry, this corresponds to $\l''_{323}\lesssim 10^{-2} \fru \mt{300\gev}^2\fru{\mtr}{\text{100 MeV}}$, which should be compared to the limits in figure \ref{Lhoriz}. In the MFV scenario, the gravitino mass has to be above $\OO(100 \text{ keV}) \times \tan^2\beta  $ (a similar bound was also studied in ref. \cite{Csaki:2011ge}, resulting in a dependence on $\tan^4\beta$). }
Further studies, especially at gravitino masses between 1 and 10 GeV, are needed. In this range, the best constraints on the RPV coupling will come from antiprotons, positrons and gamma rays. This would also imply a somewhat suppressed mediation scale for SUSY breaking, lower than $M_P$ or $M_{GUT}$, providing a suggestive hint for more new physics at intermediate energies.
In a forthcoming publication \cite{futureUS}, we are comprehensively exploring  all the different decay and detection channels, the uncertainties related to propagation and DM halo profile, as well as the full dependence on the SUSY spectrum.

\paragraph{Note Added} While the write-up of this paper was being completed, reference \cite{Savastio:2014wva} was submitted to the arXiv, which puts similar limits on gravitino DM in the MFV framework by analyzing the antiproton and $\gamma$ fluxes and has no mention of the SUSY scale. In the present work, the source of the bounds is the lack of observation of antideuterons, which is less sensitive to astrophysical uncertainties. In addition to analyzing other types of flavor symmetries and a larger range of gravitino masses, we extensively discuss sensitivity to the superpartner scale and limits from overclosure.

\section*{Acknowledgments}
%
\noindent We would like to thank Patrick Draper and Michael Dine for useful discussions. EC is supported by a NASA Graduate Research Fellowship under NASA NESSF Grant No. NNX13AO63H. JMC is supported by the NSF Graduate Research Fellowship under Grant No. (DGE-1339067).




\providecommand{\href}[2]{#2}\begingroup\raggedright\endgroup

\end{document}